# The Role of Structure in the Protein Dynamical Transition

Yunfen He and Andrea G. Markelz
University at Buffalo, SUNY, Buffalo, NY 14260, USA

*Abstract*—The protein dynamical transition is investigated as a function of protein structure using terahertz time domain spectroscopy (THz-TDS). Measurements performed for native state and denatured hen egg white lysozyme (HEWL) show that protein structure is not necessary for the dynamical transition. We find the temperature dependence follows activated behavior and there is no evidence of a fragile to strong transition. Measurements of short chain poly alanine show a dynamical transition down to penta-alanine, however no transition is observed for di-alanine or tri-alanine. These measurements demonstrate that the temperature dependence arises strictly from the interaction of the side chains with the solvent. The lack of a transition for shorter chain polypeptides may indicate the temperature dependence arises from a net ordering of the adjacent water which scales with the length of the polypeptide chain.

## I. INTRODUCTION AND BACKGROUND

Biomolecular internal motions are crucial for their function. Among the more controversial phenomena of protein dynamics is the existence of a strong temperature dependence of molecular flexibility near 200 K for proteins and polynucleotides hydrated above 30% by weight. The change in flexibility is characterized by a rapid increase in the mean square atomic displacement $<x^2>$ at a temperature $T_D \sim$ 180-220 K for a variety of proteins and nucleic acids. This rapid increase in $<x^2>$ is the so-called "dynamical transition".

A number of mechanisms have been proposed for the rapid change in flexibility. Some investigators have associated the transition with the distribution of energy barriers in the energy landscape arising from the precise 3D structure [1]. However the narrow temperature range observed for a wide variety of biomolecules and the requirement of a minimum hydration level suggests the rapid increase in flexibility arises from a change in the dynamics of the solvent in direct contact with the biological molecule, the biological water [2]. As molecular motion requires the surrounding solvent to accommodate conformational changes, any rigidity in the solvent will constrain the motion and any rapid change in the solvent dynamics will be reflected in the biomolecular dynamics. Among the proposed solvent mechanisms is a fragile to strong glass transition of the biological water [3]. This type of transition has been observed for nanoconfined water, and authors have suggested that such ordering in the adjacent water, may be realized at the biomolecular surface [4]. Another possibility is that the temperature dependence, which is often found to be Arrhenius, is simply a manifestation of activated diffusive motions either of the biological water or the surface side chains. If this is the case, then the dynamical transition should still be present even when all structure is removed.

While the dynamical transition is often measured using neutron quasi-elastic scattering, terahertz dielectric response is also sensitive to the rapid change in flexibility of the system [5]. This sensitivity arises from either the relaxational loss from picosecond diffusive motions, or low frequency structural vibrational mode absorption.

## II. MATERIALS AND METHODS

We investigate the dynamical transition as a function of protein structure using terahertz time domain spectroscopy (THz-TDS). Starting lysozyme solutions were prepared by dissolving HEWL powder (Sigma Aldrich L6876) in Trizma buffer (pH 7.0, 0.1 M) to a concentration of 200 mg/ml. Denatured solutions were prepared by adding guanidine hydrochloride (GdmHCl) to the native HEWL solutions to a GdmHCl concentration of 6M. The solutions were clear and without precipitates. While GdmCl is an excellent denaturant, it also has been demonstrated to inhibit aggregation and fibril formation. The temperature dependent THz TDS measurements of solutions follow the same procedure as discussed previously.[CPL07, APL07] THz TDS measurements were made of the pure buffer, native HEWL, denatured HEWL, and 6M GdmCl solutions.

The denaturation was characterized by UV fluorescence and circular dichroism measurements. Of some concern is the question of the extent of denaturing in the 6 M GdmHCl at such high protein concentrations. GdmHCl was chosen because of its ability to prevent aggregation and precipitation. That is for example pH denaturing can result in random coil chain to chain binding. These aggregates themselves may have some net structure, as has been found in amyloid formation. This does not appear to occur in GdmHCl solutions at sufficiently high concentration. Among the theories of the mechanism of GdmHCl denaturing is that it interrupts internal H bonding and coats the polypeptide, prevent both internal bonding and chain to chain bonding. If this is indeed the dominant mechanism then GdmHCl is a stoichiometric denaturant. As most measurements demonstrating the extend of denaturing with GdmHCl are performed on protein concentrations of 10 mg/ml or less, it is therefore possible that we did not have full denaturing at our high 200 mg/ml concentrations.

In Figure 1 we show the UV fluorescence with the characteristic shift from ~340nm for native hen egg white lysozyme (N-HEWL) to ~350nm for hen egg white lysozyme in 6M GdmHCl (D-HEWL) in the tryptophan fluorescence as the tryptophans are exposed to solvent.

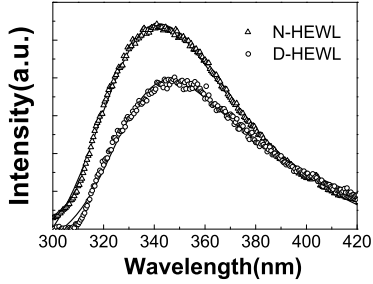

Figure 1  The fluorescent emission for N-HEWL and D-HEWL solutions (excitation at 290nm).

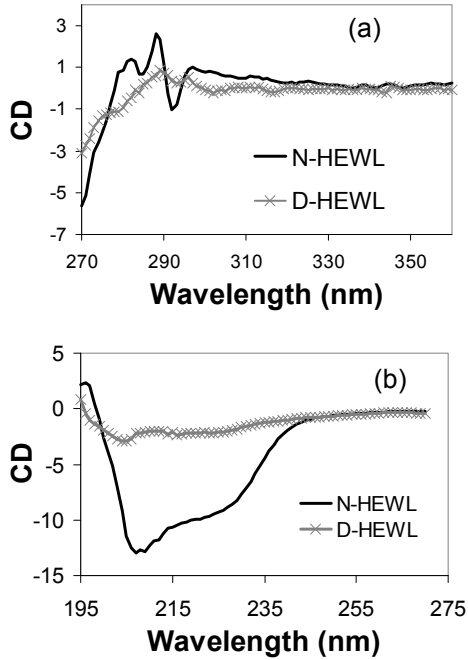

Figure 2  The tertiary structure circular CD (a) and secondary structure circular CD (b) for N-HEWL and D-HEWL solutions.

In Figure 2 we show both tertiary structure CD (a) and secondary structure CD (b).  These measurements were performed with the same quartz cell used for the THz measurements, however tertiary structure CD measurements were done with a ~20 micron thick spacer. One clearly sees from Figure 2(a) that the tertiary structure is removed in the 6 GdmHCl solutions. For the secondary structure measurements, at the 200 mg/ml concentrations the OD in this range is very high making the secondary structure characterization a challenge, even with our 20 micron spacer.  We were able to overcome this by using approximately 5 microliters pipetted on a quartz window.  The top window was allowed to sit directly on top of the bottom window without any spacer. We estimate the total thickness of the resulting cell was < 5 microns.  In Figure 2(b) we show the secondary structure CD for the N-HEWL and D-HEWL solutions.  As seen in the figure, the N-HEWL shows the expected secondary structure, while secondary structure for D-HEWL is entirely absent, even at these high protein concentrations.

## III.  RESULTS

From THz-TDS measurements for N-HEWL and D-HEWL seen in Figure 3, we observe the dynamical transition for N-HEWL, also we find the dynamical transition is still present for hen egg white lysozyme denatured in 6 M GdmHCl, indicating protein structure is not necessary for the effect. The absorption coefficient α for GdmHCl has no temperature dependence.  We also found that ln(α) vs 1/T for D-HEWL (after removing the low temperature linear contribution) appears Arrhenius temperature dependence.

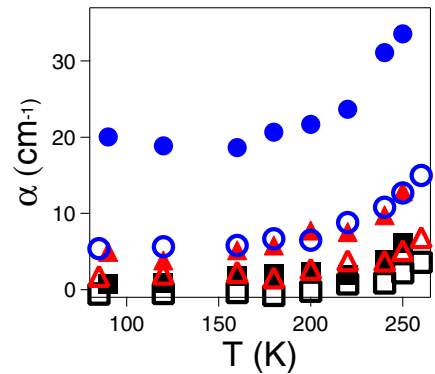

Figure 3  The temperature dependent THz absorption coefficient α for N-HEWL (unfilled symbols) and D-HEWL (filled symbols) Symbol notation: ■: 0.51THz; ▲: 0.70THz; ●: 1.00THz.

These results do not support the recent suggestion that the temperature dependence is a manifestation of a fragile to strong glass transition similar to that seen in nanoconfined water. The fragile to strong glass transition is indicated by a low temperature Arrhenius dependence and a high temperature fragile liquid temperature dependence such as a Vogel Tammann Fulcher temperature dependence ~ $\exp(-DT_o/[T-T_o])$ [6]. Our results demonstrate that the dynamical transition remains even if the nanoconfinement at the protein surface is strongly altered. Also we find Arrenhius behavior for the full temperature range. Our observed transition temperatures and Arrhenius behavior are consistent with the proposed explanation of the ~ 200 K dynamical transition arising simply from the general temperature dependence of diffusive motions, for example activated motions [7, 8]. The measurements for N-HEWL and D-HEWL establish the diffusive motions are independent of the structure in the system.

While it is still not clear that if the dynamical transition arises strictly from either the solvent or the side-chain diffusive motion.  If the origin is strictly side chain relaxations, then the temperature dependence should continue as we reduce the polymer length.

Measurements on a short chain peptide series allow us to determine the minimum chain length needed for the transition. The dynamical transition does not appear for di-alanine(A2) and tri-alanine(A3), while the transition is clearly observed for penta-alanine(A5), short chain polyalanine(A7~10) and poly-DL-alanine(PA). If the transition simply arises from diffusive behavior of the side chains, this should still be present for A2 and A3 [9, 10]. Molecular dynamics simulations have suggested that the transition originates in the biological water [11]. These results suggest that possibly the water-protein interactions that give rise to the temperature dependence have a minimum chain length dependence. The temperature dependence of the diffusive motions is the result of the polypeptide on the dynamics of biological water.

This work was supported by NSF CAREER PHY-0349256